\documentclass[12pt]{article}

\usepackage{amsmath,amsfonts,epsfig,mathrsfs,todonotes,yfonts}
\usepackage{xcolor}
\usepackage{cite}
\usepackage{stmaryrd}
\usepackage{mdframed}
\usepackage[top=2.5cm, bottom=2.5cm, left=2.75cm, right=2.75cm]{geometry} 
\usepackage{dsfont}
\numberwithin{equation}{section}
\usepackage{hyperref}
 \hypersetup{
     colorlinks=true,
     linkcolor=black,
     filecolor=blue,
     citecolor=blue,      
     urlcolor=cyan,
     }
\usepackage[most]{tcolorbox}

\newcommand{\re}[1] {(\ref{#1})}

\newcommand{\al}{\alpha}

\newcommand{\ber}{\begin{eqnarray}}
\newcommand{\eer}[1]{\label{#1}\end{eqnarray}}
\newcommand{\eero}{\end{eqnarray}}

\newcommand{\balg}{\begin{align}}
\newcommand{\ealg}{\end{align}}

\newcommand{\beq}{\begin{equation}}
\newcommand{\eeq}{\end{equation}}
\newcommand{\bea}{\begin{eqnarray}}
\newcommand{\eea}{\end{eqnarray}}

\newcommand{\nn}{\nonumber}
\newcommand{\na}{\nabla}

\newcommand{\half}{{\textstyle{\frac12}}}

\def\HollowBox #1#2{{\dimen0=#1 \advance\dimen0 by -#2
       \dimen1=#1 \advance\dimen1 by #2
        \vrule height #1 depth #2 width #2
        \vrule height 0pt depth #2 width #1
        \llap{\vrule height #1 depth -\dimen0 width \dimen1} 
       \hskip -#2
       \vrule height #1 depth #2 width #2}}
\def\BOX{\HollowBox{.100in}{.010in}}

\usepackage{fancyhdr} 
\newcommand{\auth}{\large Ulf Lindstr\"om ${}^{a,b}$\footnote{email: ulf.lindstrom@physics.uu.se}
and {\"O}zg{\"u}r Sar{\i}o\u{g}lu ${}^a$\footnote{email: sarioglu@metu.edu.tr}}
\begin{document}
\begin{flushright}
{\small UUITP-20/21}\\
\vskip 1.5 cm
\end{flushright}

\begin{center}
{\Large{\bf New currents with Killing-Yano tensors}}
\vspace{.75cm}

\auth
\end{center}
\vspace{.5cm}
\vspace{.5cm}
\centerline{${}^a${\it \small Department of Physics, Faculty of Arts and Sciences,}}
\centerline{{\it \small Middle East Technical University, 06800, Ankara, Turkey}}
\vspace{.5cm}
\centerline{${}^b${\it \small Department of Physics and Astronomy, Theoretical Physics, Uppsala University}}
\centerline{{\it \small SE-751 20 Uppsala, Sweden }}

\vspace{1cm}


\centerline{{\bf Abstract}}
\bigskip

\noindent
New relations involving the Riemann, Ricci and Einstein tensors that have to hold 
for a given geometry to admit Killing-Yano tensors are described. These relations 
are then used to introduce novel conserved ``currents'' involving such Killing-Yano tensors. 
For a particular current based on the Einstein tensor, we discuss the issue of conserved 
charges and consider implications for matter coupled to gravity. The condition on the 
background geometry to allow asymptotic conserved charges for a current introduced 
by Kastor and Traschen is found and a number of other new aspects of this current are 
commented on. In particular we show that it vanishes for rank $(D-1)$ Killing-Yano tensors 
in $D$ dimensions.
\vskip .5cm

\vspace{0.5cm}
\small

\renewcommand{\thefootnote}{\arabic{footnote}}
\setcounter{footnote}{0}

\pagebreak
\tableofcontents
\setcounter{page}{2}

\section{Introduction}\label{intro}

Killing-Yano tensors (KYTs) have long been studied in various settings. They can be thought 
of as square roots of Killing tensors with which they share some properties. In particular they 
are relevant to gravity, supergravity and string theory for finding hidden symmetries for particles 
and backgrounds, for separating variables in Hamilton-Jacobi equations and for finding the 
symmetries of the Dirac equation and its super extensions. The literature is vast and this is 
not a review, so we shall just mention some references that we have found useful in our 
present endeavor.

A general background to Killing tensors and KYTs is the nice paper \cite{Hansen}. 
A classical application to finding new supersymmetries is contained 
in ``Susy in the sky'' \cite{Gibbons:1993ap}. Relevant for string theory are the more recent 
paper \cite{DeJonghe:1996fb} and the extensive treatise \cite{Chervonyi:2015ima}. There 
are further applications in General Relativity (GR) \cite{Carter,Walker:1970un} to $G$-structures 
\cite{Papadopoulos:2007gf, Papadopoulos:2011cz}, to WZW models \cite{Lunin:2020mxj}, 
to classical mechanics \cite{Cariglia:2014dfa} and to symmetries of the Dirac operator 
\cite{Cariglia:2012ci}. A comprehensive survey of these topics, together with many more 
references, can be found in \cite{Santillan:2011sh}. Finally, supersymmetric conformal KYTs 
are discussed in \cite{Howe:2018lwu}, and partly in \cite{Kuzenko:2020www}.

We are interested in the effect of KYTs on the geometry. Part of our motivation
is purely mathematical, investigating the interplay between the properties of a generic 
rank $n$ KYT and the rest of the geometry. As a consequence, we are also 
able to construct conserved antisymmetric contravariant tensors that we 
refer to as conserved currents. Not being Noether currents, these tensors 
correspond to conserved integrals that are not in general flux integrals. They can 
nevertheless in some cases lead to conserved charges along the lines of the Abbott-Deser (AD) 
construction for a Killing vector contracted with the energy momentum tensor. Our 
setting is GR in $D$ dimensions coupled to matter. Assuming that this admits a KYT 
of rank $n$, we derive two new identities for such KYTs and use them to find new 
conserved currents. We apply our identities to several known solutions of GR and 
discuss possible conserved charges for the new currents as 
well as other constraints on the matter content. 

Our discussion is inspired by a result of Kastor and Traschen \cite{Kastor:2004jk,Kastor:2007tg}, 
who constructed a conserved current for an arbitrary rank KYT. We show how this 
KT-current\footnote{See section \ref{ktcur} for the definition of a \em{KT-current}.} in general 
splits into sums of conserved currents and how special gravitational backgrounds allow 
particular such splittings. In \cite{Kastor:2004jk,Kastor:2007tg}, it is stated that any spacetime 
that allows asymptotic KYTs will give rise to conserved charges using the KT-current. 
We find that in general there are obstructions to this, and derive a relation that the 
perturbed background geometry has to satisfy for these charges to exist. These obstructions can 
be traced back to the linearized Bianchi identities needed  for the conservation of the KT-current.

After the definition of KYTs in section \ref{KYTs}, we describe the new identities in section 
\ref{oldid} and the currents in section \ref{newcur}. In section \ref{constr}, we discuss some 
of the consequences of the existence of a KYT on the matter fields coupled to GR.
Section \ref{ktcur} contains a reformulation of the KT-current in terms of the Weyl and 
Schouten tensors. This rewrite allows us to show that the KT-current identically vanishes 
in $D=3$ for all KYTs and {for rank $n=D-1$ KYTs in $D \geq 4$.} 
It also helps us to identify new constituent currents for special dimensions and KYT ranks. 
Moreover, it contains the derivation of a condition on the geometry for a general KT-current 
to give rise to asymptotic AD charges \cite{Abbott:1981ff}. Sections \ref{constr} and \ref{ktcur} 
also contain gratifying checks on our identities for the FLRW geometry, the Kantowski-Sachs 
metric and the Kerr-Newman black hole. Section \ref{comme} deals with various special 
cases of the $n=2$ KT-current. In appendix \ref{appa}, we discuss and exclude AD charges for 
one of our new currents based on the Einstein tensor. Appendix \ref{appb} contains the 
proof that another of our currents is conserved for conformally flat geometries. 

\section{Killing-Yano tensors}\label{KYTs}
The Killing-Yano tensors generalise Killing vectors and Killing tensors 
to rank $n$ antisymmetric tensor fields with analogous properties. They can be
thought of as being the components of an $n$-form\footnote{So in $D$ dimensions, one 
has $n\leq D$.}
\( f_{a_{1} \dots a_{n}} = f_{[a_{1} \dots a_{n}]} \)
satisfying 
\ber
 \nabla_{(a} f_{b) a_{2} \dots a_{n}} = 0 ~,
\eer{Deff}
which  implies  the further properties
\ber 
\nabla_{a_{1}} f_{a_{2} \dots a_{n+1}} = \nabla_{[a_{1}} f_{a_{2} \dots a_{n+1}]} \quad
\mbox{and} \quad \nabla_{a_{1}} f^{a_{1} \dots a_{n}} = 0 \,. \label{ide0}
\eer{99} 
These can be used to derive the nontrivial identity\footnote{We use ``identity'' in the less 
strict sense where the properties of $f$ have to be taken into account.} \cite{Kastor:2004jk}
\ber
 \nabla_{a} \nabla_{b} f_{c_{1} \dots c_{n}} = (-1)^{n+1} \frac{(n+1)}{2} \,
 R^{d}\,_{a[bc_{1}} \, f_{c_{2} \dots c_{n}] d} \,, \label{ddKY}
\eer{ID1}
which generalises the analogous formula for a Killing vector 
\( \nabla_{a} \nabla_{b} f_{c} = R^{d}\,_{abc} \, f_{d} \) when $n=1$.

\section{KYT identities}\label{oldid}
Let us rewrite (\ref{ID1}) explicitly for $n=2$:
\beq
\nabla_{a} \nabla_{b} f_{cd} = - \frac{3}{2} \, R^{e}\,_{a[bc} \, f_{d] e} 
= \frac{1}{2} \, R^{e}\,_{abc} \, f_{ed} + \frac{1}{2} \, R^{e}\,_{acd} \, f_{eb}
+ \frac{1}{2} \, R^{e}\,_{adb} \, f_{ec}  \,. \label{ddKY2}
\eeq
We contract the $(a,c)$ indices in (\ref{ddKY2}). Since \( \nabla_{a} f^{ab} = 0 \), we
find that
\bea\label{nanaf}
\nabla_{a} \nabla_{b} f^{ac} & = & [\nabla_{a}, \nabla_{b}] f^{ac} =
R_{ab} \, f^{ac} + R_{ab}\,^{c}\,_{d} \, f^{ad} \nonumber \\
& = & \frac{1}{2} \, R_{ab} \, f^{ac} - \frac{1}{2} \, R^{ac} \, f_{ab}
+ \frac{1}{2} \, R_{da}\,^{c}\,_{b} \, f^{da} \,,
\eea
where the first line follows from the definition of the commutator of covariant derivatives and 
the second line from the contraction of indices on the right hand side of (\ref{ddKY2}). From the equality \re{nanaf} we find
\bea
\frac{1}{2} \left( R_{ab} \, f^{ac} + R^{ac} \, f_{ab} \right) & = &
 \frac{1}{2} \, R_{dab}\,^{c} \, f^{ad} + R_{abd}\,^{c} \, f^{ad} \nn \\
& = & \frac{1}{2} \, R_{dab}\,^{c} \, f^{da} +  R_{bda}\,^{c} \, f^{da} \,, \nn 
\eea
using \( R_{[abd]}\,^{c} = 0 \). We split the last term into two halves using
the dummy index pair $(a,d)$ and employ \( R_{[abd]}\,^{c} = 0 \) again to arrive at
\begin{equation}
\tcboxmath{ R_{ab} \, f^{ac} + R^{ac} \, f_{ab} = 0 \,. }
\label{ide1}
\end{equation}
To our knowledge the identity (\ref{ide1}) was first reported in \cite{Collinson1}, but does not 
seem to be widely known (see however \cite{Collinson2,Stephani,Ibohal}). It can alternatively 
be derived by acting on the defining property \re{Deff} with a second covariant derivative, 
considering various index combinations and applying the Ricci identity. This also leads to 
an identity between the Weyl tensor and $f$ which we omit. See \cite{Stephani,Ibohal} for details.

\subsection{Generalisation of (\ref{ide1}) for arbitrary rank $n$ KYT}\label{ident1}
We repeat the steps above for the generic rank $n$ case starting from (\ref{ID1}). 
Contracting the $(a,c_{n})$ indices gives
\begin{equation}
g^{a c_{n}} \nabla_{a} \nabla_{b} f_{c_{1} \dots c_{n}} 
= \nabla^{c_{n}} \nabla_{b} f_{c_{1} \dots c_{n}} 
= \left[ \nabla^{c_{n}}, \nabla_{b} \right] f_{c_{1} \dots c_{n}} \,,
\end{equation}
which yields
\bea
R^{d}\,_{b} \, f_{[c_{1} \dots c_{n-1}] d} 
+ (-1)^{n} (n-1) \, R_{bda[c_{1}} \, f_{c_{2} \dots c_{n-1}]}\,^{ad}
 \qquad \qquad \qquad \qquad \qquad \qquad & & \nn \\
= R^{d}\,_{[b} \, f_{c_{1} \dots c_{n-1}] d} 
+ (-1)^{n} \frac{(n-1)}{2} \, R_{ad[b c_{1}} \, f_{c_{2} \dots c_{n-1}]}\,^{ad} \,.
\label{nide1}
\eea
Since
\bea
R^{d}\,_{[b} \, f_{c_{1} \dots c_{n-1}] d} & = & \frac{1}{n} \left( R^{d}\,_{b} \, f_{[c_{1} \dots c_{n-1}] d}
+ (-1)^{n-1} (n-1) \, R^{d}\,_{[c_1} \, f_{c_{2} \dots c_{n-1}] bd} \right) \,, \nonumber \\
R_{ad[b c_{1}} \, f_{c_{2} \dots c_{n-1}]}\,^{ad} & = & \frac{1}{n} \left(
4 \, R_{bda[c_{1}} \, f_{c_{2} \dots c_{n-1}]}\,^{ad} 
+ (-1)^{n-1} (n-2) \, R_{ad[c_{1} c_{2}} \, f_{c_{3} \dots c_{n-1}]b}\,^{ad} \right) \,, \nonumber
\eea
(\ref{nide1}) can be recast as
\begin{tcolorbox}[ams align]
R^{d}\,_{b} \, f_{[c_{1} \dots c_{n-1}] d} + (-1)^{n} \, R^{d}\,_{[c_1} \, f_{c_{2} \dots c_{n-1}] bd} 
\qquad \qquad \qquad \qquad \qquad \qquad & 
\nonumber \\
+ (n-2) \left( (-1)^{n} R_{bda[c_{1}} \, f_{c_{2} \dots c_{n-1}]}\,^{ad}
        + \frac{1}{2} \, R_{ad[c_{1} c_{2}} \, f_{c_{3} \dots c_{n-1}]b}\,^{ad} \right) & = 0 \,. 
\label{new1}
\end{tcolorbox}

This is the generalisation of (\ref{ide1}) for a rank $n$ KYT, and to our knowledge,
has not been reported elsewhere in the literature\footnote{It has been pointed out to us 
by one of the referees that it might be  related to the material in subsection 3.4 
of \cite{Batista:2014fpa}. Indeed the integrability condition in \cite{Batista:2014fpa} gives 
the relation \re{ide1} for a $n=2$ KYT when traced over one set of indices. For a $n=3$ KYT 
tracing and anti-symmetrising two different index pairs recovers our \re{new1}.}. 

As a quick check, it identically reduces to (\ref{ide1}) when $n=2$. Note that when any pair 
of free indices are contracted in (\ref{new1}), one gets identically zero on the left hand side 
and there is nothing to infer from such contractions.

\subsection{A new identity}\label{pedan}
Let us go back to (\ref{ddKY2}) for a rank $n=2$ KYT. This time we differentiate, 
i.e. consider\footnote{The analogs of the steps taken here for the case of a Killing vector 
$f$, i.e. $n=1$ case, gives the well-known result that the Lie derivative of the Riemann tensor 
along the Killing vector vanishes, i.e. \( {\cal L}_{f} R_{abcd} = 0 \), which leads to 
\( {\cal L}_{f} R_{ab} = 0 \,, {\cal L}_{f} R = 0 \) and hence to  \( {\cal L}_{f} G_{ab} = 0 \).}
\beq
\nabla_{a} \left( \nabla_{b} \nabla_{c} f_{de} \right) 
- \nabla_{b} \left( \nabla_{a} \nabla_{c} f_{de} \right) = [\nabla_{a}, \nabla_{b}] \nabla_{c} f_{de} \,, 
\label{dddKY2}
\eeq
and use (\ref{ddKY2}) on the left hand side of (\ref{dddKY2}). Using the Bianchi identity
\( \nabla_{[a} \, R_{bc]d}\,^{e} = 0 \) and multiplying by an overall factor of 2, one gets
\beq
 f_{i[c} \, \nabla^{i} R_{de]ab} =  R^{i}\,_{b[cd} \, \nabla_{e]} f_{ia} +
  R^{i}\,_{a[cd} \, \nabla_{e]} f_{bi} + 2 R_{ab[c}\,^{i} \nabla_{d} f_{e]i} \,. \label{ara2}
\eeq
Contracting the $(a,e)$ indices in the latter and multiplying by an overall factor of 3 then gives
\bea
2 f_{a[d} \nabla^{a} R_{c]b} + f^{ia} \nabla_{i} R_{abcd} 
& = & 2 R_{iba[c} \nabla_{d]} f^{ia} 
+ 2 R^{a}\,_{[d} \nabla_{c]} f_{ba} + R_{iacd} \nabla_{b} f^{ia} \nn \\
& & + 4 R_{abi[d} \nabla_{c]} f^{ai}
+ 2 R^{a}\,_{b} \nabla_{c} f_{da} \, \nn \\
& = & 3 R_{abi[c} \nabla_{d]} f^{ia} + 2 R_{iab[c} \nabla_{d]} f^{ia} + 3 R^{a}\,_{[d} \nabla_{c} f_{b]a} 
\nn \\
& &+ R^{a}\,_{b} \nabla_{c} f_{da} + R_{iacd} \nabla_{b} f^{ia} \,. \nn
\eea
Finally contracting the $(b,c)$ indices in the last equality gives
\[ f_{ad} \nabla^{a} R - f^{ab} \nabla_{a} R_{db} -f^{ba} \nabla_{b} R_{da} = 0 \,, \]
which is equivalent to
\begin{equation}
\tcboxmath{ f^{ab} \nabla_{a} G_{bd} = 0 \,, \label{fdEin} }
\end{equation}
where $G_{ab}$ denotes the Einstein tensor. As far as we know, this identity has not been 
reported elsewhere.

\subsection{Generalisation of (\ref{fdEin}) for arbitrary rank $n$ KYT}
It is again worth repeating the steps taken from (\ref{dddKY2}) to (\ref{fdEin}) for a generic 
rank $n$ KYT. Starting from (\ref{ddKY}), we have, in  analogy to (\ref{dddKY2}),
\beq 
\nabla_{a} \left( \nabla_{b} \nabla_{c} f_{c_{1} \dots c_{n}} \right) 
- \nabla_{b} \left( \nabla_{a} \nabla_{c} f_{c_{1} \dots c_{n}} \right) 
= [\nabla_{a}, \nabla_{b}] \nabla_{c} f_{c_{1} \dots c_{n}} \,.
\label{dddKYn}
\eeq
Using (\ref{ddKY}), the Bianchi identity \( \nabla_{[a} \, R_{bc]d}\,^{e} = 0 \) and some
algebra, one finds
\beq
\left( \nabla_{d} R_{ab[c c_{1}} \right) f_{c_{2} \dots c_{n}]}\,^{d} 
= 2 R_{abd[c} \, \nabla_{c_{1}} \, f_{c_{2} \dots c_{n}]}\,^{d}
  + R_{bd[c c_{1}} \, \nabla_{|a|} \, f_{c_{2} \dots c_{n}]}\,^{d}
  + R_{da[c c_{1}} \, \nabla_{|b|} \, f_{c_{2} \dots c_{n}]}\,^{d}  
  \label{aran}
\eeq
analogous to (\ref{ara2}). On both sides of (\ref{aran}), if one contracts first the index pair 
$(a,c_{n})$ and then the pair $(b,c)$, one finds that the right hand side vanishes identically. 
However the left hand side yields
\begin{tcolorbox}[ams align]
(n-1) \left( \nabla^{b} \, R^{a}\,_{[c_{1}} \right) f_{c_{2} \dots c_{n-1}] ab} 
+ \frac{1}{2} \, \left( \nabla^{a} R \right) f_{a [c_{1} \dots c_{n-1}]} = 0 \,.
\label{noEin}
\end{tcolorbox}

This is the generalisation of (\ref{fdEin}) for a rank $n$ KYT, and reduces to (\ref{fdEin}) 
when $n=2$. To our knowledge, this identity is also new.

\section{New currents}\label{newcur}
Let us return to the $n=2$ case, and the associated identities (\ref{ide1}) and 
(\ref{fdEin}). The antisymmetry of the KYT and (\ref{ide1}) immediately give
\beq
G_{ab} \, f^{ac} + G^{ac} \, f_{ab} = 0 \,, \label{Einide}
\eeq
i.e. the analogous identity for the Einstein tensor. This suggests defining the ``current"\footnote{Apart
from \cite{Menekay}, the relation (\ref{Einide}) appears neither to have been considered nor used.}
\begin{equation}
\tcboxmath{ K^{ab} \equiv 2 \, G_{c}\,^{[a} \, f^{b]c} = G^{a}\,_{c} \, f^{bc} - G^{b}\,_{c} \, f^{ac} 
= 2 \, G^{a}\,_{c} \, f^{bc}  \,, \label{Eincur} }
\end{equation}
where the last equality follows from (\ref{Einide}). It is easy to see that this antisymmetric 
tensor is covariantly conserved 
\begin{equation}
\tcboxmath{\na_{a} K^{ab} = 0 \,. } \label{dEcur}
\end{equation}
This can be shown in at least two separate ways. The easier one starts by using the last equality
in (\ref{Eincur}), and employing (\ref{ide0}) and the property \( \na_{a} G^{ab} = 0 \). Alternatively,
one can use the penultimate equality in (\ref{Eincur}). This results in a total of four terms for
\( \na_{a} K^{ab} \), three of which cancel out by (\ref{ide0}) and \( \na_{a} G^{ab} = 0 \) as 
before. The remaining piece, \( ( \na_{a} \, G^{b}\,_{c}) \, f^{ac} \), does vanish due to (\ref{fdEin}).

A question that comes to mind is whether the current $K^{ab}$ \re{Eincur} can be 
used for finding new conserved Killing charges, in the sense of e.g. 
\cite{Abbott:1981ff,Kastor:2004jk}. The stakes are high because of the presence of the 
Einstein tensor, which through the Einstein field equations, naturally relates to the matter 
sources. It seems unlikely, since the current $K^{ab}$ \re{Eincur} does not have a Noether origin, 
i.e. conservation is not modulo field equations and it is not derived as a Noether current for a 
symmetry, but that fact does not exclude asymptotic charges in the sense of \cite{Abbott:1981ff}, 
(AD charges)\footnote{See subsection \ref{AppArg} for a detailed discussion.} for the KT-current. 
In appendix \ref{appa} we explicitly show and explain the absence of an asymptotic AD-charge 
for maximally symmetric spacetimes.

Perhaps naively but  naturally, one is also tempted to generalise the expression 
\re{Eincur} for $K^{ab}$ and define 
\ber
J_E^{c_1...c_n} = G^{d[c_1} \, f^{c_2...c_n]}_{~~~~~~~~d} \,. 
\label{JE}
\eer{aa}
as a possible new current. It should be noted that the covariant conservation of this expression 
requires
\ber\nn
&&\na_a J_E^{ac_2...c_n}=\frac 1 {n}\Big(\na_a G^{da}f^{c_2...c_n}_{~~~~~~~~d}+(-1)^{n+1}(n-1)\na_a G^{d[c_2}f^{c_3...c_n]a}_{~~~~~~~~~d}\\[1mm]
&&~~~~~~~~~~~~~~~~~~~~~+G^{da}\na_a f^{c_2...c_n}_{~~~~~~~~d}+(-1)^{n+1}(n-1) G^{d[c_2}\na_a f^{c_3...c_n]a}_{~~~~~~~~~d}\Big)=0~.
\eer{Eext}
Using (\ref{ide0}) and \( \na_{a} G^{ab} = 0 \), the latter becomes
\ber
\na_a J_E^{ac_2...c_n} = (-1)^{n+1} \frac{(n-1)}{n} \, 
\na_a \, G^{d[c_2} \, f^{c_3...c_n]a}_{~~~~~~~~~d} =0~.
\eer{not0}
We first observe that this expression vanishes for $n=1$. This reproduces the well-known covariant
conservation of the Killing vector current, e.g. in \cite{Kastor:2004jk}. Secondly, we note that 
\re{not0} vanishes \emph{if} 
\ber
G^{d[c_2}f^{c_3...c_n]a}_{~~~~~~~~~d}\sim G^{da}f^{[c_2c_3...c_n]}_{~~~~~~~~~~d} \,,
\eer{33}
which is true for $n=2$ according to \re{fdEin}. Nevertheless, it does not vanish for general $n$,
which can be seen from the $n$-dependent coefficient in \re{noEin}. However, it does vanish for 
special cases, such as conformally flat geometries (See appendix \ref{appb}). 

Closer scrutiny of \re{noEin} reveals that one can in fact generalise (\ref{Eincur}) for a generic
rank $n$ KYT by defining
\begin{equation}
\tcboxmath{ K^{a_{1} \dots a_{n}} \equiv R_{c}\,^{[a_{1}} \, f^{a_{2} \dots a_{n}] c} + 
\frac{(-1)^{n}}{n} \, R \, f^{a_{1} \dots a_{n}} \,, }
\label{Newcur1}
\end{equation}
that is covariantly conserved, \( \nabla_{a_{1}} K^{a_{1} \dots a_{n}} = 0 \), and reduces 
to (\ref{Eincur}) for $n=2$. 

\section{Constraints on matter sources from \re{Eincur} and \re{dEcur}}\label{constr}
In this section we restrict our attention to the consequences of \re{Eincur} and \re{dEcur} on
continuous matter distributions that are described by a stress-energy tensor $T_{ab}$,
which acts as a source in Einstein's field equations. To keep the discussion concise, we 
only consider the stress tensors of a perfect fluid and  of an electromagnetic field.

\subsection{The perfect fluid}\label{perflu}
The stress tensor of a perfect fluid is given by
\beq
T_{a b} = \rho u_{a} u_{b} + p \left(g_{a b} + u_{a} u_{b} \right) \,,
\eeq
where \( u^{a} \) is a unit timelike 4-velocity of the fluid with \( u^{a} u_{a} = -1 \) and
the functions $p$ and $\rho$, respectively, denote the pressure and the mass-density of 
the fluid.
The stress tensor satisfies the equations of motion
\beq
\nabla^{a} T_{a b} = 0 \,, \label{delTab}
\eeq
which yields
\bea
u^{a} \nabla_{a} \rho + (\rho + p) \nabla^{a} u_{a} & = & 0 \,, \label{Wa1} \\
(p + \rho) u^{a} \nabla_{a} u_{b} + \left( g_{a b} + u_{a} u_{b} \right) \nabla^{a} p & = & 0  \,. \label{Wa2}
\eea
If the spacetime of interest admits a KYT of rank $n=2$, then \re{dEcur}, or equivalently 
\re{fdEin} which becomes \( f^{ab} \nabla_{a} T_{bc} = 0 \), imposes yet another set of 
conditions in analogy to \re{Wa1} and \re{Wa2} above. These read
\bea
f^{ab} u_{b} \nabla_{a} \rho + (\rho + p) f^{ab} \nabla_{a} u_{b} & = & 0 \,, \label{LS1} \\
(p + \rho) f^{ab} u_{b} \nabla_{a} u_{c} 
+ \left( g_{b c} + u_{b} u_{c} \right) f^{ab} \nabla_{a} p & = & 0 \,. \label{LS2}
\eea

The new identities \re{LS1} and \re{LS2} can be checked by using e.g. the Robertson-Walker 
metric written as
\beq
ds^2 = -dt^2 + a^{2}(t) \big( dr^2 + b^{2}(r) ( d\theta^2 + \sin^{2}{\theta} \, d\varphi^{2} ) \big) \,, 
\label{rwmet4}
\eeq
where \( b(r) \equiv \sin{r}, r, \sinh{r} \) corresponding to the three spatial -- spherical, Euclidean,
hyperboloidal, respectively -- geometries. This metric admits four independent rank $n=2$
KYTs\cite{Ibohal}, the components of which read
\bea
f_{(1)\, \theta r} = 2 a^{3} b \sin{\varphi} \,, \;
f_{(1)\, \varphi r} = a^{3} b \cos{\varphi} \sin{2\theta} \,, \;
f_{(1)\, \theta \varphi} = 2 a^{3} b^{2} b^{\prime} \cos{\varphi} \sin^{2}{\theta} \,; \nn \\
f_{(2)\, r \theta} = 2 a^{3} b \cos{\varphi} \,, \;
f_{(2)\, \varphi r} = a^{3} b \sin{\varphi} \sin{2\theta} \,, \;
f_{(2)\, \theta \varphi} = 2 a^{3} b^{2} b^{\prime} \sin{\varphi} \sin^{2}{\theta} \,; \nn \\
f_{(3)\, r \varphi} = 2 a^{3} b \sin^{2}{\theta} \,, \;
f_{(3)\, \theta \varphi} = a^{3} b^{2} b^{\prime} \sin{2\theta} \,; \nn \\
f_{(4)\, \theta \varphi} = 2 a^{3} b^{3} \sin{\theta} \,. \label{Ibos}
\eea
Here we have omitted the arguments of the metric functions $a$ and $b$, and used a prime over
$b$ to indicate differentiation with respect to $r$. One can show separately for each KYT 
\re{Ibos} that \re{LS1} and \re{LS2} (as well as \re{Wa1} and \re{Wa2}, of course) are satisfied
for the Robertson-Walker metric.

As for another example, one can consider the Kantowski-Sachs metric in $D=4$:
\beq
ds^2 = -dt^2 + X^{2}(t) dr^2 + Y^{2}(t) ( d\theta^2 + \sin^{2}{\theta} \, d\varphi^{2} ) \,. 
\label{ksmet4}
\eeq
This is a solution of the Einstein field equations for dust and admits the rank $n=2$
KYT\cite{Ibohal} with a single component
\beq
f_{\theta \varphi} = 2 Y^{3}(t) \sin{\theta} \,. \label{Iboks}
\eeq
It follows easily that \re{LS1} and \re{LS2} (as well as \re{Wa1} and \re{Wa2}, of course) are 
satisfied for the Kantowski-Sachs metric.

\subsection{The electromagnetic field}\label{MaxField}
The electromagnetic stress tensor is given by 
\beq
T_{a b} = F_{a c} F_{b}\,^{c} - \frac{1}{4} g_{a b} F_{d e} F^{d e} \,.
\eeq
From the Einstein field equations, one must again have that (\ref{delTab}) is satisfied. 
Using \( \nabla_{[a} F_{b c]} = 0 \) carefully, this yields
\beq
\nabla^{a} T_{a b} = (\nabla^{a} F_{a c}) F_{b}\,^{c} = 0 \,. \label{cond1}
\eeq
If Maxwell's equations admit a current, then they read
\beq
\nabla^{a} F_{a b} = j_{b} \,,
\eeq
and \re{cond1} can be thought of as \( F^{bc} j_{c} = 0 \), a non-trivial requirement to be satisfied
by the components of the current. For a nontrivial solution for the current $j_c$, the ``coefficients"
$F^{bc}$ must be such that \( {\rm det} (F^{bc}) = 0 \)\footnote{In $D=4$, one has  
\( {\rm det} (F^{a b}) \sim (F_{a b} \,^{*}F^{a b})^2 \), of course.}. Put in another way, one must have 
\( \nabla^{a} F_{a b} = 0 \) provided  \( {\rm det} (F^{bc}) \neq 0 \).

If the spacetime of interest admits a KYT of rank $n=2$, then \re{dEcur}, or equivalently 
\re{fdEin} which becomes \( f^{ab} \nabla_{a} T_{bc} = 0 \), imposes 
\beq 
( f^{ab} \nabla_{a} F_{bd}) F_{c}\,^{d} 
+ \frac{3}{2} F^{bd} \nabla_{a} \big( f^{a}\,_{[b} \, F_{cd]} \big) = 0 \,. \label{cond2}
\eeq

The celebrated Kerr-Newman solution in $D=4$ is an example for which the new 
identities put forward can be checked. The metric and the vector potential are given by
\bea
d s^{2} & = & - \left( \frac{\Delta - a^{2} \sin^{2} \theta}{\Sigma} \right) d t^{2} 
  - \frac{2 a \sin^{2}\theta \left( r^{2} + a^{2} - \Delta \right)}{\Sigma} \, d t \, d \phi \nn \\
& & + \left( \frac{\left( r^{2} + a^{2} \right)^{2} - \Delta \, a^{2} \sin^{2} \theta}{\Sigma} \right) 
\sin^{2}\theta \, d \phi^{2} + \frac{\Sigma}{\Delta} \, d r^{2} + \Sigma \, d\theta^{2} \,,
\label{KerrNewman} \\
A_{a} \, dx^{a} & = & - \frac{q r}{\Sigma} \left( dt - a \sin^{2}\theta \, d\phi \right) \,,
\eea
where
\beq
\Sigma = r^{2} + a^{2} \cos^{2}\theta \qquad \mbox{and} \qquad
\Delta = r^{2} + a^{2} +q^{2} -2 M r \,. \label{SigDel}
\eeq
One has \( G_{ab} = 2 T_{ab} \) and \( \nabla^{a} F_{a b} = 0 \) here, with 
\( F_{ab} = 2 \partial_{[a} A_{b]} \) as usual.
Kerr-Newman metric shares the same rank $n=2$ KYT with the Kerr metric, i.e.
\re{KerrNewman} with $q=0$. Its components explicitly read
 \beq
 f_{rt} = a \, \cos{\theta} \,, \;\;
 f_{t \theta} = a r \sin{\theta} \,, \;\; 
 f_{\phi r} = a^2 \cos{\theta} \, \sin^{2} \theta \,, \;\; 
 f_{\theta \phi} = r (r^{2} + a^{2}) \sin{\theta} \,. \label{KY2KN}
 \eeq
 One can show explicitly that the identities \re{ide1}, \re{fdEin}, \re{dEcur} (together with \re{Eincur})
 and \re{cond2} are all nontrivially satisfied for the Kerr-Newman metric.
 
\section{The KT-current}\label{ktcur}
In this section we discuss under what condition conservation of a general rank $n$ 
KT-current gives rise to asymptotically conserved charges, rewrite the KT-current in terms 
of the Weyl and Schouten tensors and show that this current vanishes for rank $n=D-1$ 
KYTs in $D$ dimensions.

In \cite{Kastor:2004jk}, a covariantly conserved current\footnote{We shall refer to (\ref{KTcur}) 
as the \emph{KT-current} henceforth.} was constructed 
\ber
j^{a_{1} \dots a_{n}} = - \frac{(n-1)}{4} \, R^{[a_{1} a_{2}}\,_{b c} \, f^{a_{3} \dots a_{n}] b c}
+ (-1)^{n+1} \, R_{c}\,^{[a_{1}} \, f^{a_{2} \dots a_{n}] c} - \frac{1}{2 n} \, R \, f^{a_{1} \dots a_{n}} \,,
\eer{KTcur}
with \( \nabla_{a_{1}} j^{a_{1} \dots a_{n}} = 0 \), for a spacetime that admits a rank 
$n$ KYT. To show the conservation of $j^{a_{1} \dots a_{n}}$ the following Bianchi identities 
are needed:
\ber
&&\na_{[a}R_{bc]}{}^{de} = 0 \,, \quad 
\na_{a} R_{bc}{}^{da} + 2 \na_{[b} R_{c]}^{~d} = 0\,, \quad 
\na_{a} R^{a}{}_{b} - \half \na_b R = 0 \,.
\eer{Bian}
A look at the newly found current \re{Newcur1} shows that one can in fact split
the KT-current into two separately covariantly conserved pieces. To see this, introduce
\beq
\tcboxmath{ \tilde{K}^{a_{1} \dots a_{n}} = 
- \frac{(n-1)}{4} \, R^{[a_{1} a_{2}}\,_{b c} \, f^{a_{3} \dots a_{n}] b c}
 + \frac{1}{2 n} \, R \, f^{a_{1} \dots a_{n}} \,, }
\label{Newcur2}
\eeq
with \( \nabla_{a_{1}} \tilde{K}^{a_{1} \dots a_{n}} = 0 \) and write the KT-current as \cite{Acik:2008qe}
\beq
\tcboxmath{
j^{a_{1} \dots a_{n}} = \tilde{K}^{a_{1} \dots a_{n}} + (-1)^{n-1} K^{a_{1} \dots a_{n}} }
\,. \label{KTsplit}
\eeq

\subsection{AD charges for the KT-current}\label{AppArg}
A covariantly conserved antisymmetric rank $n$ tensor field is equivalent 
to a co-closed $n$-form. By the extension of the Poincar\'e lemma to the exterior 
co-derivative this means that it is equal to the co-derivative of an $(n+1)$-form in an 
open set, under quite general conditions on this set. In what follows we apply this fact 
to the background geometry to construct conserved charges for linearized currents.

In\cite{Kastor:2004jk,Cebeci:2006mc}, the existence of asymptotic charges 
based on the KT-current was shown for asymptotically flat and asymptotically AdS 
geometries. The method is a generalisation of the idea of employing asymptotic 
Killing vectors \cite{Abbott:1981ff} to define the corresponding conserved charges.

We first treat the current based on a rank-2 KYT. So consider a $D$-dimensional 
spacetime $\bar{g}_{ab}$, which is often referred to as ``the background spacetime" with 
a completely antisymmetric rank-2 KYT $\bar{f}_{a b}$ satisfying 
\beq
{\bar{\nabla}}_{a} \, \bar{f}_{b c} + {\bar{\nabla}}_{b} \, \bar{f}_{a c} = 0 \,. \label{nyaneq}
\eeq
Now the spacetime $g_{ab}$ whose \emph{new} Killing-Yano charge(s) we are after does 
not necessarily have to admit exact KYTs. We assume that the metric 
$g_{ab}$ can be asymptotically split into a background plus a deviation as
\beq
g_{ab} \equiv \bar{g}_{ab} + h_{ab} \quad \mbox{so that} \quad 
g^{ab} = \bar{g}^{ab} - h^{ab} + {\cal O}(h^2) \,, \label{metspl}
\eeq 
where $ h^{ab}=\bar{g}^{ac}h_{cd}~\!\bar{g}^{db}$.
In what follows, all indices are raised and lowered with the generic background metric 
$\bar{g}_{ab}$, e.g. \( h \equiv \bar g^{ab} h_{bc} \) and \( \bar{\BOX} \equiv 
{\bar{\nabla}}^{c} \, {\bar{\nabla}}_{c} \). To ${\cal O}(h)$ his leads to the following linearized 
curvature, Ricci tensor and curvature scalar:
\ber
( R_{ab}{}^{cd} )_{L} & = & \bar{R}_{abe}{}^{[c} h^{d]e} + 2 \, \bar{\na}_{[a} \bar{\na}^{[d} h_{b]}{}^{c]} 
\,, \label{RiemL} \\[1mm]
( R^{a}{}_{b} )_{L} & = & \frac{1}{2} \left( \bar{\na}^{c} \bar{\na}^{a} h_{bc} + 
\bar{\na}_{c} \bar{\na}_{b} h^{ac} - \bar{\na}^{a} \bar{\na}_{b} h - \bar{\BOX} h^{a}{}_{b} \right)
- h^{ac} \bar{R}_{bc} \,,
\\[1mm]
R_{L} & = & {\bar{\nabla}}_{a} {\bar{\nabla}}_{b} h^{ab} - \bar{\BOX} h - h^{ab} \bar{R}_{ab} \,.
\eer{Lin}
To see if the {\em linearised} KT-current is conserved, we shall need the following versions 
of the identities \re{Bian} that hold modulo terms of ${\cal O}(h^2)$ and higher:
\bea
&& \bar\na_{[a}(R_{bc]}{}^{de})_L + (\Gamma_{[a})_L \cdot \big(\bar R_{bc]}{}^{de} \big) = 0 \,, 
\nn \\[1mm]
&& \bar\na_{a}(R_{bc}{}^{da})_L + 2 \bar\na_{[b}(R_{c]}^{~d})_L + (\Gamma_{a})_L \cdot 
\big( \bar R_{bc}{}^{da} \big) + 2 (\Gamma_{[b})_L \cdot 
\big( \bar R_{c]}^{~d} \big) = 0 \,, \label{LBian} \\[1mm]
&& \bar\na_{a}(R^{a}{}_{b})_L - \half \bar\na_b R_L + (\Gamma_{a})_L \cdot 
\big( \bar R^{a}{}_{b} \big) = 0 \,.  \nn 
\eea
Here $(\Gamma_{a})\cdot $ denotes the usual action of a connection on a tensor as exemplified by
\ber
(\Gamma_{a}) \cdot ( T^{b}{}_{c}) = \Gamma^{b}{}_{ae} T^{e}{}_{c} - \Gamma^{e}{}_{ac} T^{b}{}_{e} \,.
\eer{gamdot}
The linearized connection is
\ber
(\Gamma_{~ab}^{c})_L = \half \bar g^{ce} \big( \bar\na_a h_{be} + \bar\na_b h_{ae} 
- \bar\na_e h_{ab}\big) \,.
\eer{LinG}
Note that for flat or maximally symmetric backgrounds, the relations \re{LBian} become the same 
as \re{Bian} with all curvature objects replaced by their linearized counterparts. It is this form that 
is needed for background conservation of the linearised current. We shall also need the assumption 
that $g_{ab}$ asymptotically admits KYTs due to this splitting and that $h_{ab}$ vanishes sufficiently 
fast at the hypersurface of interest $\Sigma$ (see \re{dtty} below) which is used for defining the 
charges. When the linearized connection terms in \re{LBian} vanish, the current  $j^{ab}$ is 
background covariantly conserved, i.e. \( \bar{\nabla}_{a} (j^{ac})_{L} = 0 \). Since the current 
is antisymmetric, the covariant conservation is expected to give rise to an ordinary conservation 
law via
\beq
\bar{\nabla}_{a} (j^{ac})_{L} = \frac{1}{\sqrt{|\bar{g}|}} \,
\partial_{a} \big( \sqrt{|\bar{g}|} \, (j^{ac})_{L} \big) = 0 \,. \label{conslaw}
\eeq
From this we infer as usual that the integral
\ber
\int d^{D-1} x \, {\sqrt{|\bar{g}|} \, (j^{0b}})_{L} 
\eer{bb} 
is constant. In \cite{Kastor:2004jk,Cebeci:2006mc} the latter is turned into a flux integral over a 
$(D-3)$-dimensional hypersurface\footnote{For a rank $n$ KYT, the analogous step involves 
an integral over a hypersurface of dimension $D-1-n$.} by further invoking the Stokes' theorem: 
The {\em crucial} step is the determination of the potential for the current, as described in 
the beginning of this section. We thus need to express $(j^{ac})_{L}$ as the divergence of a 
completely antisymmetric rank-3 tensor \( (j^{ac})_{L} = {\bar{\nabla}}_{d} \, \bar{\ell}^{acd} \,, \)
where \( \bar{\ell}^{acd} = \bar{\ell}^{[acd]} \). Then, up to a trivial normalization, the conserved 
``charge" can be obtained by
\beq
Q^{ac} \sim \int_{\Sigma} \, dS_{i} \, \sqrt{|\bar{\gamma}|} \, \bar{\ell}^{aci} \,, \label{dtty}
\eeq
where $i$ ranges over the $(D-3)$-dimensional hypersurface $\Sigma$ and $\gamma$
is the induced metric on $\Sigma$. 

The asymptotic charges for the KT-current were given in \cite{Kastor:2004jk} for an arbitrary 
rank $n$ KYT in an asymptotically flat background and in \cite{Cebeci:2006mc} for an arbitrary 
rank $n$ KYT in a maximally symmetric background. Their existence again rests on the KT-current  
being expressible as the covariant divergence of an $(n+1)$-form. Since the 
construction of such a potential is nontrivial, here we complement this discussion 
by deriving a condition that the background has to satisfy for such an $(n+1)$-form to exist.

Following \cite{Kastor:2004jk}, the general rank KT-current can be written as
\ber
j^{a_{1} \dots a_{n}} = N_{n} \, \delta^{a_{1} \dots a_{n} d_{1} d_{2}}_{b_{1} \dots b_{n} c_{1} c_{2}}
\, f^{b_{1} \dots b_{n}} \, R_{d_{1}d_{2}}{}^{c_{1}c_{2}} \,, 
\eer{KTcom}
where
\( \delta^{a_{1} \dots a_{m}}_{b_{1} \dots b_{m}} = 
\delta^{[a_{1}}_{b_{1}} \cdots \delta^{a_{m}]}_{b_{m}} \)
is totally antisymmetric in all up and down indices, and
\ber
N_{n} = - \frac{(n+1)(n+2)}{4n} \,.
\eer{cc}
As explained above, we are only interested in the linearized part of \re{KTcom} and find 
\ber
(j^{a_{1} \dots a_{n}})_{L} = N_{n} \, 
\delta^{a_{1} \dots a_{n} d_{1} d_{2}}_{b_{1} \dots b_{n} c_{1} c_{2}} \, 
\bar{f}^{b_{1} \dots b_{n}} \, (R_{d_{1}d_{2}}{}^{c_{1}c_{2}})_{L} \,.
\eer{KTComL}
In terms of the linearized Riemann tensor in \re{RiemL}, the current may be written
\ber
(j^{a_{1} \dots a_{n}})_{L} = N_{n} \, 
\delta^{a_{1} \dots a_{n} d_{1} d_{2}}_{b_{1} \dots b_{n} c_{1} c_{2}} \, \bar{f}^{b_{1} \dots b_{n}} \,
\big( \bar{R}_{d_{1} d_{2} e}{}^{[c_{1}} \, h^{c_{2}] e} 
+ 2 \, \bar{\na}_{d_{1}} \bar{\na}^{c_{2}} h_{d_{2}}{}^{c_{1}} \big) \,.
\eer{dd}
In \cite{Kastor:2004jk} it is shown that, for a flat background, this may be written as 
\ber
(j^{a_{1} \dots a_{n}})_{L}  = \bar{\nabla}_{e} \, \bar{\ell}^{e a_{1} \dots a_{n}}
\eer{ee}
where the $(n+1)$-form \( \bar{\ell}^{e a_{1} \dots a_{n}} = \bar{\ell}^{[e a_{1} \dots a_{n}]} \) is
\ber
\bar{\ell}^{e a_{1} \dots a_{n}} = 2 N_{n} \, 
\delta^{a_{1} \dots a_{n} e d_{2}}_{b_{1} \dots b_{n} c_{1} c_{2}} \,
\bar{f}^{b_{1} \dots b_{n}} \, \bar{\nabla}^{c_2} \, h_{d_{2}}{}^{c_{1}}
- \frac{1}{2n} \Big( h \, \bar{\nabla}^{e} \, \bar{f}^{a_{1} \dots a_{n}}
- (n+1) \, h^{d_{2} [e} \, \bar{\nabla}_{d_{2}} \, \bar{f}^{a_{1} \dots a_{n}]} \Big) \,.
\eer{ell}
Similar manipulations as in \cite{Kastor:2004jk} give the following result for the general 
case\footnote{Note that there are no additional curvature terms generated in the process.}
\ber
(j^{a_{1} \dots a_{n}})_{L} = \bar{\nabla}_{e} \, \bar{\ell}^{e a_{1} \dots a_{n}}
+ N_{n} \Big( \bar{f}^{[a_{1} \dots a_{n}} \, \bar{R}_{c_{1} c_{2} e}{}^{c_{1}} \, h^{c_{2}]e}
+ 2 \, h_{c_{2}}{}^{[c_{1}} \, \bar{\nabla}^{c_{2}} \, 
\bar{\nabla}_{c_{1}}  \, \bar{f}^{a_{1} \dots a_{n}]} \Big)
\eer{Rdiff}
with $\bar{\ell}$ as in \re{ell}.

Using \re{ID1} and the explicit antisymmetrisation, vanishing of the terms in parenthesis 
\re{Rdiff} can be expressed in terms of the background curvature as\footnote{When $n=1$,
\re{ADcnd} simply reads \( h \bar{R}^{ab} \bar{f}_{b} - h^{bc} \bar{R}_{bc} \bar{f}^{a} = 0 \).}
\ber
\tcboxmath{
 \bar{f}^{[a_{1} \dots a_{n}} \, \bar{R}_{c_{1} c_{2} e}{}^{c_{1}} \, h^{c_{2}]e}
  + 2 (-1)^{n} h_{c_{2}}{}^{[c_{2}} \bar{R}_{e}{}^{c_{1}}{}_{c_{1}}{}^{a_{1}} \, 
 \bar{f}^{a_{2} \dots a_{n}] e} = 0 \,.}
\eer{ADcnd}

For the KT construction of  asymptotic charges, the condition \re{ADcnd} has to hold.
It is clearly fulfilled for the flat case which leads to the results in \cite{Kastor:2004jk}.
For a  maximally symmetric background 
\[ \bar{R}_{abcd} = \frac{2 \Lambda}{(D-1)(D-2)} \, 
(\bar{g}_{ac} \, \bar{g}_{bd} - \bar{g}_{ad} \, \bar{g}_{bc}) \,, \;\;
\bar{R}_{ab} = \frac{2 \Lambda}{(D-2)} \, \bar{g}_{ab} \,, \;\;
\bar{R} = \frac{2 \Lambda D}{(D-2)} \,, \;\;
\bar{G}_{ab} = - \Lambda \, \bar{g}_{ab} \,. \]
 \re{ADcnd} is also fulfilled and leads to the results in \cite{Cebeci:2006mc}. This agrees 
 with the known cases where the linearized Bianchi identities \re{LBian} ensure conservation 
 of the KT-current.
 
Using the expansion of the Riemann tensor in terms of the Weyl and Schouten 
tensors, given in \re{WtoR} below, \re{ADcnd} may alternatively be written as
\begin{tcolorbox}[ams align]
\bar{f}^{[a_{1} \dots a_{n}} \, \bar{C}_{c_{1} c_{2} e}{}^{c_{1}} \, h^{c_{2}]e}
  + 2 (-1)^{n} h_{c_{2}}{}^{[c_{2}} \bar{C}_{e}{}^{c_{1}}{}_{c_{1}}{}^{a_{1}} \, 
 \bar{f}^{a_{2} \dots a_{n}] e} 
 \qquad \qquad \qquad & \nonumber \\[1mm] 
 - \textstyle{\frac{2 (D-(n+1))}{n+2}} \, 
  \left( \bar{f}^{[a_{1} \dots a_{n}} \, h^{d_{1}] d_{2}} \, \bar{S}_{d_{1} d_{2}}
 + h_{d_{1}}{}^{[a_{1}} \, \bar{S}_{d_{2}}{}^{d_{1}} \, \bar{f}^{|d_{2}| a_{2} \dots a_{n}]} \right) & = 0 \,.
\label{ADcnd3}
\end{tcolorbox}
This expression may be further simplified using the traceless property of the Weyl tensor.
 
\subsection{KT-current in terms of the Weyl and Schouten tensors}
It is also instructive to rewrite the KT-current \re{KTcur} using the 
decomposition of the Riemann tensor in terms of the Weyl tensor $C$ and the Schouten
tensor $S$
\bea
S_{ab} & \equiv & \frac{1}{(D-2)} \left( R_{ab} - \frac{1}{2(D-1)} R \, g_{ab} \right) \,, \nonumber \\
R_{ab} & = & (D-2) S_{ab} + S \, g_{ab} \quad \mbox{with} \quad S \equiv g^{ab} S_{ab} \;\;
\mbox{so that} \;\; R = 2(D-1) S \,, \nonumber \\
R^{ab}{}_{cd} & = & C^{ab}{}_{cd} + 4 \delta^{[a}{}_{[c} \, S^{b]}{}_{d]} \quad \mbox{and} \quad
G_{ab} = (D-2) \left( S_{ab} - S \, g_{ab} \right) \,. \label{WtoR}
\eea
These let one express \re{KTcur} alternatively as
\bea
j^{a_{1} \dots a_{n}} & = & - \frac{(n-1)}{4} \, C^{[a_{1} a_{2}}\,_{b c} \, f^{a_{3} \dots a_{n}] b c}
+ (-1)^{n-1} \, \left( \frac{D-(n+1)}{D-2} \right) \, R_{c}\,^{[a_{1}} \, f^{a_{2} \dots a_{n}] c} \nonumber \\
& & + \left( \frac{n-1}{2(D-1)(D-2)} - \frac{1}{2n} \right) \, R \, f^{a_{1} \dots a_{n}} \,, \nonumber \\
& = & - \frac{(n-1)}{4} \, C^{[a_{1} a_{2}}\,_{b c} \, f^{a_{3} \dots a_{n}] b c} \nonumber \\
& & + (D-(n+1)) \, \left( (-1)^{n-1} \, S_{c}\,^{[a_{1}} \, f^{a_{2} \dots a_{n}] c}
- \frac{1}{n} \, S \, f^{a_{1} \dots a_{n}} \right) \,.
\eea
The latter equality shows that when the rank $n=D-1$, the KT-current \re{KTcur} reduces to
\beq
j^{a_{1} \dots a_{n}} \Big|_{n=D-1} =  
- \frac{(D-2)}{4} \, C^{[a_{1} a_{2}}\,_{b c} \, f^{a_{3} \dots a_{n}] b c} \,. \label{spec}
\eeq
Note also that since the Weyl tensor vanishes identically in $D=3$, so does the whole
KT-current $j^{ab}$ for $n=2$. Moreover, when $D=4$ one has a special current 
for a rank $n=3$ KYT from \re{spec} 
\beq
j^{a_{1} a_{2} a_{3}} =  
- \frac{1}{2} \, C^{[a_{1} a_{2}}\,_{b c} \, f^{a_{3}] b c} \,. \label{spec4}
\eeq
In fact one can show that this also vanishes and thus the KT-current does not exist in this 
case either. The Hodge dual of a KYT is a closed conformal Killing tensor (KT) \cite{Frolov:2008jr}. 
In particular this means that a rank $n=D-1$ KYT is dual to a closed conformal Killing vector, (defined in \re{CKY} below),
as discussed in \cite{Batista:2014uja}. We thus first dualize the $n=3$ KYT to a closed conformal Killing vector $\tilde{f}_a$ 
(defined in \re{CKY} below)  to write
\beq
\tilde{f}_{a} = \frac{\sqrt{|g|}}{3! } \, \epsilon_{a}{}^{bcd} \, {f}_{bcd} \quad 
\Rightarrow \quad  {f}_{bcd} = \tilde{f}_{a} \, \epsilon^{a}{}_{bcd}
\label{flrw1}
\eeq
satisfying 
\beq
\nabla_{a} \tilde{f}_{b} = \frac{1}{4} (\nabla_{c} \tilde{f}^{c}) \, g_{ab} \,.
 \label{flrw2}
\eeq
Dualizing also $j^{a_{1} a_{2} a_{3}}$, we may then write the relation \re{spec4} up to 
some signs and factors as
\ber
\epsilon_{d a_{1} a_{2} a_{3}} j^{a_{1} a_{2} a_{3}} \sim 
\epsilon_{d a_{1} a_{2} a_{3}} \, C^{a_{1} a_{2}}{}_{bc} \, \epsilon^{a_{3} bce} \, \tilde{f}_{e} \,.
\eer{dual1}
Using the formula for contracting one index on the Levi-Civita symbol and the traceless
property of the Weyl tensor then shows that the right hand side vanishes, and thus that 
\( j^{a_{1} a_{2} a_{3}} = 0 \). This can also be seen, perhaps more directly,  from the fact that 
\ber
C_{abcd} \, \tilde{f}^{d} = 0
\eer{dual2}
in $D=4$ when $\tilde{f}^d$ satisfies \re{flrw1}, see e.g. \cite{Stephani}. We have 
\ber
C^{a_{1} a_{2}}{}_{bc} \, \epsilon^{a_{3} bce} \, \tilde{f}_{e} = 2 \, \tilde{f}_{e} \, 
C^{a_{1} a_{2} a_{3} e} \, \star = 2 \, \star C^{a_{1} a_{2} a_{3} e} \, \tilde{f}_{e} = 0 \,,
\eer{zilch}
where we used a relation between the right and left duals of the Weyl tensor 
(see, e.g. \cite{Portillo}) and the last equality follows by \re{dual2}.

The condition \re{dual2} implies that either $\tilde{f}_{e}$ is a null vector or the space 
is conformally flat. It is gratifying to see that at least for the conformally flat case
the existence of the charge condition \re{ADcnd3} also vanishes. 

Clearly the argument leading from \re{spec4} to \re{zilch} holds equally well for a KT-current 
based on a rank $n=D-1$ KYT in $D$ dimensions, so that such a KT-current also has to vanish.

\section{Comments on the KT and related currents}\label{comme}
The KT current has many interesting special cases for particular geometries. We also 
found that a number of conserved ``currents''  related  to the KT current can be defined. 
In this section we summarize these cases for completeness and collect their interrelations 
in a table.

We start by reproducing the $n=2$ KT-current \re{KTcur}, for convenience:
\bea
-4 j^{ab} & = & R^{abcd} \, f_{cd} + 4 R_{c}\,^{[a} \, f^{b]c} + R \, f^{ab} \,. \label{KTcur2} \\
 & = & C^{abcd} \, f_{cd} + 2(D-3) \left( 2 S^{c[a} \, f^{b]}\,_{c} + S f^{ab} \right) \,, 
\label{alt1}
\eea
It is interesting to note that the expression multiplying $(D-3)$ in \re{alt1} 
\ber
J_{(1)}^{ab} \equiv 2 f^{c[a} \, S_{c}{}^{b]} + f^{ab} S
\eer{98}
is conserved for certain geometries. We have
\ber
\na_a J_{(1)}^{ab} = f^{ca} \, \na_{a} S^{b}{}_{c} 
\eer{J1con}
which vanishes when $S_{ab}$ is a Codazzi tensor, i.e. a symmetric 2-tensor whose 
covariant derivative is also symmetric
\ber
\na_a S_{c}{}^{b} = \na_c S_{a}{}^{b} \,.
\eer{ff}
The Weyl-Schouten theorem \cite{Weyl,Schouten} states that:
\begin{quote}
{A Riemannian manifold of dimension $D$ with $D \geq 3$ is conformally 
flat if and only if the Schouten tensor is a Codazzi tensor for $D =3$, or the Weyl tensor 
vanishes for $D > 3$.}
\end{quote}
Hence we need a conformally flat metric in $D=3$ for $J_{(1)}^{ab}$ to be conserved. 
For higher dimensions, we note that a {\em metric $g$ has a harmonic Weyl tensor
\ber
\na_aC_{cd}^{~~ab}=0~,
\eer{gg}
if and only if its Schouten tensor is a Codazzi tensor}. In this case we see 
from \re{alt1} that $j^{ab}$ is the sum of two independently conserved currents, 
one proportional to $J_{(1)}^{ab}$ and a new current
\ber
J_{(2)}^{ab} \equiv f^{cd} \, C_{cd}^{~~ab}~,
\eer{hatJ}
according to ($D>3$)
\ber
-4 j^{ab} = J_{(2)}^{ab} + 2 (D-3) J_{(1)}^{ab}~.
\eer{hh}

A rank $n$ conformal Killing-Yano tensor ($CKYT$) $\hat f$ obeys
\ber
\na_{b}\hat f_{a_1...a_{n}}=\na_{[b}\hat f_{a_1...a_{n}]}+ng_{b[a_1}\bar f_{a_2...a_n]}
\eer{CKY}
with
\ber
\bar f_{a_2 \dots a_n} \equiv \frac 1{D-n+1} \na_{b} f^b_{~a_2...a_{n}} \, .
\eer{ckytdef}
When the first term in \re{CKY} vanishes, the tensor is called a closed conformal Killing-Yano tensor (CCKYT). A differential form is a KYT if, and only if, its Hodge dual is a CCKYT.

The current $J_{(2)}^{ab}$ in \re{hatJ} can be extended to involve a conformal Yano 2-form $\hat f$.
When acting on by the covariant derivative
\ber
\na_a \big( \hat f_{bc} \, C^{cdaf} \big) =  \big( \na_{[a} \hat f_{bc]}
+ 2 g_{a[b} \bar f_{c]} \big) C^{cdaf}+ \hat f_{bc}\na_a C^{cdaf}\,,
\eer{ii}
the first term vanishes due to the anti-symmetrization of $\na\hat f$ which imposes the first Bianchi identity on $C$, the second vanishes since 
$C$ is trace-free and the third since the Weyl tensor is harmonic.

It may also be of interest to consider a metric $g$ with a harmonic Riemann tensor
\ber
\na_a R_{cd}^{~~ab} = 0~.
\eer{Rharm}
This {\em requires the Ricci tensor to be a Codazzi tensor, instead of the Schouten tensor}:
\ber
\na_a R_{bc} = \na_b R_{ac}~.
\eer{RicCod}
Returning to the form \re{KTcur2} for the current $J^{ab}$ we note that 
\ber
J_{(3)}^{ab} \equiv f^{cd} \, R_{cd}^{~~ab}
\eer{jj}
satisfies 
\ber
\na_a {J}^{ab}_{(3)} = g_{ae} \na^{[e} \, f^{cd]} \, R_{cd}^{~~ab}
+ f^{cd} \na_a R_{cd}^{~~ab} = 0 \,,
\eer{kk}
where the first term vanishes due to the first Bianchi identity and the second due to \re{Rharm}. 
Since the full current $j^{ab}$ is conserved, we realize that writing
\ber
J_{(4)}^{ab} \equiv j^{ab} - J^{ab}_{(3)}
\eer{ll}
yields, in analogy to the harmonic Weyl tensor case, {\em a third} current, which must be 
conserved, 
\ber
\na_a J_{(4)}^{ab} = 0~,
\eer{mm}
due to \re{RicCod}, which may also be explicitly verified.

\begin{table}[htp]
\begin{center}
\renewcommand{\arraystretch}{2} 
\begin{tabular}{|c|c|c|}
\hline
Current& Cons. Conditions&Relation to the KT-current $j^{ab}$\\
\hline
\hline
$J^{ab}_{(1)} = 2 f^{c[a} \, S^{~b]}_c + f^{ab} S$&$S_{ab}$ Codazzi &$-4j^{ab}=C^{abcd} \, f_{cd} +2(D-3)J_{(1)}^{ab}$\\
\hline
$J_{(2)}^{ab} \equiv f^{cd} \, C_{cd}^{~~ab}$&Weyl harmonic&$-4j^{ab}=J_{(2)}^{ab} +2(D-3)J_{(1)}^{ab}$\\
\hline
$J_{(3)}^{ab} \equiv f^{cd} \, R_{cd}^{~~ab}$& Riemann harmonic&$-4j^{ab}=J_{(3)}^{ab}+4 R_{c}\,^{[a} \, f^{b]c} + R \, f^{ab}$\\
\hline
$J_{(4)}^{ab}=4 R_{c}\,^{[a} \, f^{b]c} + R \, f^{ab}$&$R_{ab}$ Codazzi   &$ -4j^{ab}=J_{(3)}^{ab}+{J_{(4)}}^{ab}$\\
\hline

\end{tabular}
\end{center}
\label{default}
\caption{Relations between various currents in section \ref{comme}.}
\end{table}

\section{Conclusions and comments}
In this paper we have presented new identities for KYTs and shown how they may be 
used to find new conserved currents. These currents are all of the Kastor-Traschen type, 
i.e. not Noether currents. As shown in \cite{Kastor:2004jk,Cebeci:2006mc}, such currents 
may nevertheless lead to asymptotically conserved charges of AD type. We found a condition 
for such conserved charges to exist for the KT-current. We also displayed the linearized 
form of the Bianchi identities and pointed out that only for certain backgrounds do they 
directly lead to background conserved linearized KT-currents. An interesting question is if 
there are other backgrounds and/or modifications of the current that allow such conservation 
using these linearized identities.

For our current $K^{ab}$, based on the Einstein tensor, we investigated this possibility too 
and showed that it does not give an AD charge for a maximally symmetric space time (see 
appendix \ref{appa}). There are however many more cases, both currents and backgrounds,
that should be studied.

It is particularly interesting to note that we were able to find new conserved currents for $n>2$ 
KY forms. These should be relevant for higher dimensional solutions to Einstein's equation.

There are several directions into which the present efforts may be extended: Treating conformal 
KYTs as we touched upon in the text. Extending the geometry to allow for torsion which will 
introduce modified Killing-Yano equations as in e.g. \cite{Papadopoulos:2011cz}. This opens up 
for supersymmetric extensions, such as discussed in \cite{Howe:2018lwu}.
\bigskip

\noindent{\bf Acknowledgments}\\
{\"O}.S. would like to thank D.O. Devecio\u{g}lu for help with xAct in the early stages of this 
work. The research of U.L. is supported in part by the 2236 Co-Funded 
Scheme2 (CoCirculation2) of T\"UB{\.I}TAK (Project No:120C067)\footnote{\tiny However 
the entire responsibility for the publication is ours. The financial support received from 
T\"UB{\.I}TAK does not mean that the content of the publication is approved in a scientific 
sense by T\"UB{\.I}TAK.}. We are grateful to Dr. Vojtech Witzany for asking us a question 
that led to the clarification around \re{LBian} and to an anonymous referee for helping
us to further refine the related arguments.
\bigskip
\eject

\centerline{\large \bf APPENDICES}
\appendix
\section{No AD charge for $K^{ab}$ in maximally symmetric spacetimes.}\label{appa}
This appendix serves as an illustration for the method of deriving AD charges and its 
limitations. We adapt and apply the arguments given in subsection \re{AppArg} to the 
currents $K^{a}=G^a_{~c}f^c$ and $K^{ab}=G^{[a}_{~c}f^{b]c}$ \re{Eincur} for the 
maximally symmetric and flat backgrounds. We show explicitly that only the first can 
be used in defining new conserved quantities as done in \cite{Kastor:2007tg,Cebeci:2006mc} 
for the KT-current. 

So one starts with a $D$-dimensional background $\bar{g}_{ab}$ admitting a rank-2 
KYT $\bar{f}_{a b}$ satisfying \re{nyaneq}. For such a maximally symmetric spacetime, one has
\[ \bar{R}_{abcd} = \frac{2 \Lambda}{(D-1)(D-2)} \, 
(\bar{g}_{ac} \, \bar{g}_{bd} - \bar{g}_{ad} \, \bar{g}_{bc}) \,, \;\;
\bar{R}_{ab} = \frac{2 \Lambda}{(D-2)} \, \bar{g}_{ab} \,, \;\;
\bar{R} = \frac{2 \Lambda D}{(D-2)} \,, \;\;
\bar{G}_{ab} = - \Lambda \, \bar{g}_{ab} \,. \]
Then one finds the following which are frequently used in the ensuing calculations:
\bea
{\bar{\nabla}}_{a} \, \bar{f}^{ab} & = & 0 \,, \qquad
{\bar{\nabla}}_{a} \, \bar{f}_{bc} = {\bar{\nabla}}_{b} \, \bar{f}_{ca} 
= {\bar{\nabla}}_{c} \, \bar{f}_{ab}  \,, \label{ide2} \\
{\bar{\nabla}}_{a} \, {\bar{\nabla}}_{b} \, \bar{f}_{cd} & = & 
\frac{2 \Lambda}{(D-1)(D-2)} (\bar{g}_{ab} \, \bar{f}_{dc} + 
\bar{g}_{ac} \, \bar{f}_{bd} + \bar{g}_{ad} \, \bar{f}_{cb}) \,, \label{ide3} \\
\bar{\BOX} \, \bar{f}_{ab} & = & \frac{2 \Lambda}{(D-1)} \, \bar{f}_{ba} \,, \qquad
{\bar{\nabla}}^{a} \, {\bar{\nabla}}_{b} \, \bar{f}_{ac} =
\frac{2 \Lambda}{(D-1)} \, \bar{f}_{bc} \,. \label{ide5}
\eea
The ``linearized" version of $K^{a}=G^a_{~c}f^c$ is background covariantly conserved, 
i.e. \( \bar{\nabla}_{a} (K^{a})_{L} = 0 \). It should therefore have a potential 
$(K^{a})_{L} =\bar{\na}_d \bar{\ell}^{[da]}$ according to the general argument.

Keeping in mind that all indices are raised and lowered with the maximally 
symmetric background metric $\bar{g}_{ab}$, one finds that the linearized Ricci 
tensor and the Ricci scalar read\footnote{These easily follow by 
adapting \re{Lin} accordingly to a maximally symmetric background.}
\bea
(R_{ab})_{L} & = & \frac{1}{2} \left( {\bar{\nabla}}^{c} \, {\bar{\nabla}}_{b} \, h_{ac} 
+ {\bar{\nabla}}^{c} \, {\bar{\nabla}}_{a} \, h_{bc} - \bar{\BOX} \, h_{ab} 
- {\bar{\nabla}}_{a} \, {\bar{\nabla}}_{b} \, h \right) \,, \\
R_{L} & = & {\bar{\nabla}}^{c} \, {\bar{\nabla}}^{d} \, h_{cd} - \bar{\BOX} \, h 
- \frac{2 \Lambda}{(D-2)} h \,.
\eea
These further give
\bea
(G^{a}\,_{b})_{L} & = & \frac{1}{2} \left( {\bar{\nabla}}^{c} \, {\bar{\nabla}}^{a} \, h_{bc} 
+ {\bar{\nabla}}_{c} \, {\bar{\nabla}}_{b} \, h^{ac} - \bar{\BOX} \, h^{a}\,_{b} 
- {\bar{\nabla}}^{a} \, {\bar{\nabla}}_{b} \, h \right) \nonumber \\
& & - \frac{1}{2} \, \delta^{a}\,_{b} \, \left( {\bar{\nabla}}_{c} \, {\bar{\nabla}}_{d} \, h^{cd} 
- \bar{\BOX} \, h - \frac{2 \Lambda}{(D-2)} h \right) - \frac{2 \Lambda}{(D-2)} \, h^{a}\,_{b} \,.
\eea
With this linearized Einstein tensor the current can be rearranged to 
\ber
 (K^{a})_{L}=(G^{a}\,_{b})_{L} \bar{f}^b=\bar{\na}_d \bar{\ell}^{[da]}~.
\eer{}
Arguments analogous to those given in the discussion surrounding \re{conslaw} can now 
be repeated, replacing \( (j^{ac})_{L} \) with \( (K^{a})_{L} \), and lead to a conserved 
charge as in \re{dtty}.

The ``linearized" version of the antisymmetric ``current"\footnote{As shown in section \ref{newcur}, 
\( \nabla_{a} K^{ac} = 0 \) if the spacetime $g_{ab}$ admits a KYT $f_{ab}$ itself.}
\( K^{ac} = 2 \, G_{b}\,^{[a} \, f^{c]b} \),
\beq
(K^{ac})_{L} = - (G^{a}\,_{b})_{L} \, \bar{f}^{bc} + (G^{c}\,_{b})_{L} \, \bar{f}^{ba} \label{kcurl}
\eeq
can be similarly treated leading to
\bea
(K^{ac})_{L} & = & 3 \bar{\nabla}_{d} \left\{ \bar{f}^{b[a} \bar{\nabla}^{c} h^{d]}\,_{b} 
+ h_{b}\,^{[d} \bar{\nabla}^{c} \bar{f}^{a]b} + \frac{1}{2} \bar{f}^{[dc} \bar{\nabla}^{a]} h \right\} 
\nonumber \\
& & + \bar{\nabla}_{d} \left\{ \bar{f}^{b d} \bar{\nabla}^{[c} h^{a]}\,_{b} 
+ h_{b}\,^{d} \bar{\nabla}^{c} \bar{f}^{b a} + \frac{1}{2} \bar{f}^{c a} \bar{\nabla}^{d} h 
\right. \nonumber \\
& & \qquad \quad \left. 
+ \bar{f}^{b[a} \bar{\nabla}_{b} h^{c] d} + \bar{f}^{a c} \bar{\nabla}_{b} h^{b d} 
- h^{b d} \bar{\nabla}_{b} \bar{f}^{a c} + h \bar{\nabla}^{d} \bar{f}^{a c} \right\} 
\nonumber \\
& & + \frac{4 \Lambda}{(D-1)(D-2)} \left( h \bar{f}^{c a} + 2 h_{b}\,^{[c} \bar{f}^{a]b} \right) \,.
\label{lacd}
\eea
The first line is in the desired structure but the remaining parts of (\ref{lacd}) do not fulfill the 
requirements of a proper $\bar{\ell}$. This is so even when one takes $\Lambda \to 0$, the
same choice as in \cite{Kastor:2004jk}, to work in an asymptotically flat background. This shows
that the current $K^{ab}$ \re{Eincur} does not admit the construction of an $AD$-charge.

In retrospect the reason for this is clear. When defining \( K^{ac} = 2 \, G_{b}\,^{[a} \, f^{c]b} \), 
we needed to use \re{ide1}:
\begin{equation}
{ R_{ab} \, f^{ac} + R^{ac} \, f_{ab} = 0 \,. }
\end{equation}
When $ f^{ac}=\bar f^{ac}$ is a background KYT, as in \re{kcurl}, this holds with the 
background Ricci tensor 
$\bar R_{ab}$, and the {\em background current} \( \bar K^{ac} = 2 \, \bar G_{b}\,^{[a} \, \bar f^{c]b} \) 
is background conserved. However this will not in general be the case for the {\em linearized current} 
\( (K^{ac})_L = 2 \, (G_{b}\,^{[a})_L\, \bar f^{c]b} \) since 
\begin{equation}
 (R_{ab})_L \, \bar f^{ac} + (R^{ac})_L \, \bar f_{ab} \ne 0 \,.  
\end{equation}
Indeed the background covariant divergence of \re{lacd} is easily seen to be nonvanishing.

\section{Conservation of $J_E$ in conformally flat geometries}\label{appb}
In this section, we show that conformal flatness in fact guarantees the conservation of the
current $J_E$ in \re{JE} for an arbitrary rank $n$ KYT.
Using
\ber
R_{abc_{2}d} \, f_{c_{3} \dots c_{n}}\,^{ad} = - \frac{1}{2} \, R_{adbc_{2}} \, f_{c_{3} \dots c_{n}}\,^{ad} 
\,,
\eer{}
we rewrite \re{nide1} as
\ber\nn
&& \frac{(-1)^{n+1}}{2} \Big( 2 R^d_{~[b} \, f_{c_{2} \dots c_{n}]d}
+ (-1)^{n+1} (n-1) \, R^{da}\,_{[bc_2}f_{c_{3} \dots c_{n}]ad} \Big) \\ [1mm]
& & \qquad\qquad\qquad\qquad\qquad
= (-1)^{n+1} \, R^d_{~b} \, f_{ c_{2} \dots c_{n}d} -
\frac{(n-1)}{2} \, R_{adb[c_2} \, f_{c_{3} \dots c_{n}]}\,^{ad} \,.
\eer{317}
Using \re{WtoR}  gives
\ber\nn
& \frac{(-1)^{n+1}}{2} \Big( 2 R^d_{~[b} \, f_{c_{2} \dots c_{n}]d}
+ (-1)^{n+1} (n-1) \, C^{da}_{~~[bc_2} \, f_{c_{3} \dots c_{n}]ad} 
- 4 (-1)^{n+1} (n-1) S^d_{~[c_2} \, f_{c_{3} \dots c_{n}b]d} \Big) \\ [1mm] \nn
& = (-1)^{n+1} R^d_{~b} \, f_{c_{2} \dots c_{n}d}
- \frac{(n-1)}{2} C^{ad}_{~~b[c_{2}} \, f_{c_{3} \dots c_{n}]ad}
- (n-1) \big( S^d_{~[c_{2}} \,f_{c_{3} \dots c_{n}]bd}
- S^d_{~b} \, f_{c_{3} \dots c_{n} c_{2} d}\big) \,. \\[1mm]
\eer{318} 
For vanishing Weyl tensor and ignoring the metric terms in $S_{ab}$ \re{318} becomes
\ber
A R^d_{~[b} \, f_{c_{2} \dots c_{n}]d} = B R^d_{~b} \, f_{c_{2} \dots c_{n}d}
- C R^d_{~[c_2} \, f_{c_{3} \dots c_{n}]bd} \,,
\eer{319}
where
\[ A \equiv (-1)^{n+1} \big( 1+2 \al (1-n) \big) \,, \quad
B \equiv (-1)^{n+1} \big( 1+ \al (1-n) \big) \,, \quad
C \equiv - \al (1-n) \,, \]
with \( \al = 1/(D-2) \) depending on the dimension $D$ of the spacetime according to \re{WtoR}. Observing that
\ber
n R^d_{~[b} \, f_{c_{2} \dots c_{n}]d} = 
R^d_{~b} \, f_{c_{2} \dots c_{n}d} 
- (-1)^{n+1} (n-1) R^d_{~[c_2} \, f_{c_{3} \dots c_{n}]bd} \,,
\eer{oo}
\re{319} can be rewritten as
\ber
&&(Bn-A) R^d_{~b} \, f_{c_2...c_nd} = (C n + (-1)^{n+1} A ) R^d_{~[c_2} \, f_{c_3...c_n]bd}~,\\[1mm]\nn
&&\iff (-1)^{n} \big( \al(n-1)n-n+1+2\al(1-n) \big) R^d_{~b} \, f_{~ c_2...c_nd} \\ [1mm] \nn
&&~~~~~~~~ = (n-1) \big(-\al n+2\al +1\big) R^d_{~[c_2} \, f_{c_3...c_n]bd} \\ [1mm] \nn
&&\iff (-1)^{n+1} R^d_{~b} \, f_{c_2...c_nd} = R^d_{~[c_2} \, f_{c_3...c_n]bd} \\ [1mm]
&&\iff R^d_{~b} \, f_{c_2...c_nd} = R^d_{~[c_2} \, f_{bc_3...c_n]d}~.
\eer{322}
This leads to \re{33} which guarantees the conservation of $J_E$, provided that the metric 
terms in the Schouten tensor also work out. However, from \re{318} this requires
\ber\nn
-2 \delta^d_{~[c_2} \, f_{c_3...c_nb]d} =
- \delta^d_{~[c_2} \, f_{c_3...c_n]bd} + \delta^d_{~b} \, f_{c_3...c_nc_2d}~,
\eer{pp}
which indeed holds. So this proves that conformal flatness guarantees the conservation of the
current $J_E$ \re{JE} for an arbitrary rank $n$ KYT.

\end{document}